# Time evolution of the microwave second-order response of YBaCuO powder

A. Agliolo Gallitto, I. Ciccarello, C. Coronnello and M. Li Vigni

*INFM and Dipartimento di Scienze Fisiche e Astronomiche, via Archirafi 36,*

*I-90123 Palermo, Italy*

**Abstract**

Transient effects in the microwave second-order response of YBaCuO powder are investigated. The time evolution of the second harmonic signal has been measured for about 300 s after the sample had been exposed to variations of the DC magnetic field. We show that in different time scales the transient response has different origin. In the time scale of milliseconds the transient response of samples in the critical state is ascribable to processes of flux redistribution induced by the switching on/off of the microwave field. At longer times, the time evolution of the second harmonic signal can be ascribed to motion of fluxons induced by the variation of the DC magnetic field. In particular, diffusive motion of fluxons determines the response in the first 10 seconds after the stop of the magnetic field variation; magnetic relaxation over the surface barrier determines the response in the time scale of minutes.

**Corresponding Author**

Prof. Maria Li Vigni
Dip.to di Scienze Fisiche e Astronomiche, via Archirafi 36, I-90123 Palermo (Italy)
Tel: +39 091 6234208; Fax: +39 091 6162461; E-mail: livigni@fisica.unipa.it

# 1. Introduction

It is well known that superconducting materials, especially high-$T_c$ superconductors, when exposed to intense electromagnetic fields, show an enhanced nonlinear response [1-12]. In particular, in the presence of DC magnetic fields, the magnetization vector contains Fourier components oscillating at the second-harmonic (SH) frequency of the driving field. Several mechanisms responsible for the SH emission have been highlighted, their effectiveness depends on the temperature and the external magnetic field intensity [1, 5, 6, 9]. When a superconducting sample is exposed to both a DC magnetic field higher than the lower critical field and an intense microwave (mw) magnetic field, the SH emission arises from the motion of fluxons induced by the mw field. It has been shown that, because of the fluxon lattice rigidity, superconductors in the critical state operate a "rectification" process of the mw field, which leads to SH emission [6].

Recently, two effects have been highlighted in the second-order response of superconductors in the critical state, exposed to a static magnetic field and an intense pulsed mw magnetic field [13-15]. The SH signal exhibits time evolution during the time in which the mw pulse endures. Moreover, a magnetic hysteresis has been detected: both the intensity and the time evolution of the SH signal depend on the way the static magnetic field has been reached, i.e. at increasing or decreasing values. It has been observed that, after the static field has developed a critical state, the SH response to the very first mw pulse exhibits a peculiar time evolution, different from that relative to the response to the following pulses [14]. When a train of several mw pulses is used, after the first 4-5 pulses, a steady-state response is detected, showing a characteristic time evolution, which depends on the investigated sample, the input power level and the way in which the static field has been reached [15]. These transient effects are characterized by response times of the order of milliseconds; they have been ascribed to processes of fluxon redistribution induced by the switching on/off of the mw field.

In this paper we discuss novel transient effects in the microwave second-order response of superconductors, which occur in the time scale of seconds and minutes. We will show that in the different time scales the SH transient response may arise from different sources. In particular, in the time scale of seconds or minutes the time evolution of the SH signal can be ascribed to motion of fluxons induced by the variations of the DC magnetic field. Two different processes have been highlighted: diffusive motion of fluxons determines the response in the first ~ 10 seconds; magnetic relaxation over the surface barrier determines the response in the time scale of minutes.

**2. Experimental procedure and sample**

Measurements have been performed in a sample consisting of ~ 10 mg of powdered YBaCuO, with $T_c \approx 86$ K. The sample is sealed in a Plexiglas holder and placed in a bimodal cavity, resonating at two angular frequencies $\omega$ and $2\omega$, with $\omega/2\pi \approx 3$ GHz, in a region in which the mw magnetic fields $\boldsymbol{H}(\omega)$ and $\boldsymbol{H}(2\omega)$ are maximal and parallel to each other. The $\omega$-mode of the cavity is fed by a train of mw pulses, with maximal peak power ~ 50 W (input peak power of the order of 10 W brings on mw fields of the order of 10 Oe in the region of the cavity in which the sample is located). The pulse width and pulse repetition rate can be properly regulated to allow performing different kinds of measurements. The SH signals radiated by the sample are detected by a superheterodyne receiver. The cavity is placed between the expansions of an electromagnet, which generates static magnetic fields, $H_0$, up to $\approx 10$ kOe. All measurements have been performed, with $\boldsymbol{H}_0 \parallel \boldsymbol{H}(\omega) \parallel \boldsymbol{H}(2\omega)$ in wide ranges of temperature, DC magnetic field and input power level.

We will report on experimental results of SH emission, obtained following different types of measurements. In all the cases, before performing any measurement the sample was zero-field cooled down to the desired temperature value; the DC field was increased up to 10 kOe and then decreased down to the residual field of the electromagnet. This procedure has a double purpose: first, all the superconducting grains should have been exposed to fields greater than the full penetration field $H^*$; second, SH signals arising from nonlinear processes due to the presence of weak links are suppressed by the trapped flux.

## 3. Experimental results

A kind of measurements concerns the detection of the time evolution of the SH signal during the time in which the mw pulse endures. To this purpose, the DC magnetic field is set to a prefixed value with the mw field off and after few seconds a train of mw pulses, with pulse width ~ 1 ms, is applied. Fig. 1 shows the time evolution of the SH signal at $T = 4.2$ K during the first four pulses of the mw field, switched on after the DC magnetic field had reached the value of 3 kOe, at increasing (circles) and decreasing (squares) fields. The signal decays noticeably within the first pulse and to a minor extent within the subsequent pulses.

A peculiarity of the time evolution of the SH signal in the time scale of milliseconds, after the DC field has reached a fixed value, is the dependence of the decay rate on the input power level. As an example, in Fig. 2 we show the time evolution of the SH signal during the very first mw pulse, after the field has reached the value $H_0 = 3$ kOe at decreasing fields, for different input power levels. As one can see, the slope of the SH-vs-$t$ curve increases on increasing the input power level.

Exposing the sample to several mw sequential pulses, a steady state response is reached, in which no detectable signal decay within the pulse width is observed. However, detecting the SH signal in a time scale of the order of seconds or minutes we found a further decrease of its intensity. In order to investigate the transient effects in this time scale we have simultaneously exposed the sample to trains of mw pulses, with pulse width $5\mu s$ and pulse repetition rate 200 Hz, and a varying DC magnetic field. By measuring the intensity of the SH signal while the DC magnetic field is increasing, or decreasing, at a constant rate we have obtained the field dependence of the SH signal at $t = 0$; by stopping the magnetic-field sweep at fixed values, we have measured the time evolution of the signal for ~ 300 s, with sampling time ~ 0.5 s, from the instant in which each $H_0$ value had been fixed.

Fig. 3 shows the magnetic field dependence of SH signal intensity at $T = 4.2$ K, obtained while $H_0$ was varying from ~ 300 Oe to ~ 10 kOe, and back, at a rate of 30 Oe/s. The arrows indicate the path followed during measurements. As one can see, the SH signal intensity is higher at decreasing than at increasing fields in the whole range of fields investigated; enhanced dips are detected soon after the field-sweep direction is reversed. Measurements performed at different input power levels have shown that the shape, as well as the amplitude, of the magnetic-hysteresis loop is independent of the input power.

In Fig. 4 we report the time evolution of the SH signal, at $T = 4.2$ K, during ~ 300 s, from the instant in which $H_0$ has been set at fixed values, reached at increasing and decreasing fields. We wish to remark that these measurements afford additional information to that obtained following the procedure previously described. Indeed, the results shown in Figs. 1 and 2 have been obtained switching on the mw field few seconds after the DC magnetic field had reached a fixed value; therefore, they most likely refer to the SH response of samples in the critical state. On the contrary, by detecting the signal very soon after the $H_0$ sweep has been stopped to a fixed value we get the SH response during the time in which the critical state is developing.

The time evolution of the signal shown in Fig. 4 is characterized by two different regimes: during the first ~ 10 s the SH-vs-$t$ curves are well described by an exponential law; at longer times, the signal shows a logarithmic decay. The experimental results have been fitted by the following expressions:

$$\text{SH} = A + B \exp(-t/\tau) \qquad 0 < t < 10 \text{ s} \qquad (1)$$

and

$$\text{SH} = C\,[1-D\log(t/t_0)] \qquad t > 10 \text{ s} \tag{2}$$

with $t_0 = 10$ s.

The lines in Fig. 4 are the ones which best-fit the experimental data.

Fig. 5 shows the field dependence of the best-fit parameters $\tau$ and $D$, defined by Eqs. (1) and (2), respectively. Circles refer to the results obtained after a positive variation of the DC magnetic field, squares refer to those obtained after a negative field variation; lines are guides for eyes. In the following we indicate with $\tau_{H\uparrow}$ and $D_{H\uparrow}$ the values of the parameters obtained for increasing fields and $\tau_{H\downarrow}$ and $D_{H\downarrow}$ those obtained for decreasing fields. As one can see from Fig. 5, both $\tau_{H\uparrow}$ and $\tau_{H\downarrow}$ are roughly independent of $H_0$ and they take on equal values, within the experimental uncertainty. On the contrary, $D$ increases on increasing $H_0$ and, further, $D_{H\uparrow} < D_{H\downarrow}$.

By elaborating the experimental data of the SH time evolution, obtained at different values of the DC magnetic field, we have seen that no significant variations of both the shape and the amplitude of the hysteresis loop occur in the time scale investigated.

From measurements performed sweeping the DC magnetic field at different variation rates, we have deduced that the intensity of the SH signal observed during the field sweep and, consequently, the shape of the hysteresis loop are independent of the field-sweeping rate. A weak variation is, instead, observed in the time evolution of the signal. Fig. 6 shows the values of $\tau$ and $D$, deduced by fitting the data for $H_0 = 3$ kOe, as a function of the field-sweeping rate. Both $\tau$ and $D$ are independent of the field-sweeping rate as long as it takes on values higher than 10 Oe/s; for smaller values, both parameters increase on decreasing the field-sweeping-rate. Similar results have been obtained at different values of $H_0$.

In order to investigate the role that the mw field intensity plays in the long-time relaxation of the SH signal, we have measured the time evolution of the SH signal at different input power levels. As expected, the intensity of the SH signal increases on increasing the input power; nevertheless, the parameters characteristic of the long-time evolution of the SH signal, after the sample has been exposed to a variation of the DC magnetic field, do not depend on the input power.

Since for times longer than 10 s the variation of the signal intensity is very small, in order to investigate the time evolution of the SH signal is necessary to maintain very stable the temperature of the sample. So, we report results at $T = 4.2$ K, where the temperature is controlled by the liquid helium bath. We have also performed measurements at different temperatures, from 4.2 K to $T_c$. In some cases the parameters can only be determined with large uncertainty; however, we have found that up to 0.9 $T_c$ the peculiarities of both the hysteretic behaviour and the time evolution of the SH signal are roughly independent of the temperature. In particular, $\tau$ and $D$ take on values of the same order of those obtained at $T = 4.2$ K. Instead, for $T > 0.9$ $T_c$ some peculiarities of the SH signal change. In particular, for temperatures smaller than 0.9 $T_c$ the magnetic hysteresis of the SH signal is present in the whole range of magnetic fields investigated (0 ÷ 10 kOe), on further increasing the temperature the range of $H_0$ in which the hysteresis is present shrinks and its height reduces; eventually, at $T \geq 0.97$ $T_c$ no hysteresis is observed for whatever values of $H_0$. Measurements of the time evolution have shown that in the range 0.9 $T_c$ ÷ 0.97 $T_c$ the SH signal evolves only for the $H_0$ values at which the hysteresis is observed; when the hysteresis is missing the signal intensity does not change on elapsing the time. For $T > 0.97$ $T_c$, neither hysteresis nor decay is observed, regardless of the $H_0$ value. Therefore, the experimental results point out a strong correlation between hysteresis and decay of the SH signal.

**4. Discussion**

Non-linear magnetization of type II superconductors has been studied for the first time by Bean. The Bean model [16] applies to superconductors in the critical state and it is based on the hypothesis that the critical current does not depend on the magnetic field; furthermore, it tacitly assumes that the fluxon lattice follows adiabatically the em field variations. On these hypotheses, only odd harmonic emission is expected. However, it has been reported that superconductors in the critical state, exposed to intense pulsed mw fields, exhibit odd as well as even harmonic emission [6]. Ciccarello *et al.* [6] have elaborated a phenomenological theory, based on the Bean model, in which the additional hypothesis that superconductors in the critical state operate a "rectification" process is put forward. The hypothesis arises from the fact that, because of the rigidity of the fluxon lattice, the induction field inside the sample does not follow adiabatically the variations of high frequency fields, except when the variations involve motion of fluxons in the surface layers of the sample. It has been assumed that for the critical state developed by increasing fields, the induction flux does not vary at all during the positive half-period of the mw field, while it does during the negative one. The opposite occurs when decreasing fields develop the critical state. On this hypothesis, the response of the sample will be uneven during the wave period, with consequent odd as well as even harmonic emission. From this model, it is expected that superconductors in a critical state à la Bean radiate stationary SH signals independent of both the intensity and the sweep direction of the DC field.

More recently, it has been suggested a generalization of the model of Ciccarello *et al.*, by taking into account non-equilibrium processes, which may arise when superconductors in the critical state are exposed to high frequency fields [13]. Unlike the hypothesis of Ref. [6], it has been suggested that, even when the fluxon lattice cannot follow adiabatically the mw field variations, some fluxons actually move in the sample, with a characteristic rate related to fluxon redistribution processes. Such processes originate by the fact that when a superconductor in the critical state is exposed to fast variations of the magnetic field, the fluxons cannot quickly redistribute to a new critical state. So, a surface current arises, which induces diffusive processes in the fluxon lattice. This idea can justify the time decay of the SH signal in the time scale of milliseconds reported in Figs. 1 and 2. Indeed, the increase of the signal relaxation rate on increasing the input power level corroborates the hypothesis that the mw pulses induce the processes responsible for the signal decay. Nevertheless, these processes cannot justify the long-time decay because both $\tau$ and $D$ do not depend on the input power.

In order to further understand the role of the mw field in the processes responsible for the long-time decay of the SH signal, we have varied the DC magnetic field up to fixed values in the absence of the mw field. At $t = t_0$, when the desired $H_0$ value has been reached, we have switched on the mw field delayed of $\Delta t$ with respect to $t_0$. Fig. 7 shows the time evolution of the SH signal at $H_0 = 9$ kOe, reached at increasing fields, for different $\Delta t$ values. The results at $\Delta t = 0$ (open symbols) are the same as those obtained sweeping the DC field in the presence of the mw field. Full symbols differ each other for the different $\Delta t$ values. These results suggest that the long-time relaxation of the SH signal is not ascribable to processes induced by the mw field. On the contrary, they point out that the intensity of the SH signal, at whatever time it is detected, depends on the time elapsed from $t_0$. From these results we can infer that the intensity of the SH signal is related to the fluxon configuration, which evolves with time.

The time evolution of the SH signal could arise from relaxation of the critical state toward the thermal equilibrium state through flux creep; however, several experimental evidences disagree with this hypothesis. Firstly, at $T = 4.2$ K the decay by thermal processes is expected to be much slower than the one we observe in SH emission. On the other hand, decay times should be strongly affected by the temperature, at variance with our experimental results. Furthermore, flux-creep processes do not justify the different rates of the signal decay observed after increasing and decreasing field sweeps for times of the order of minutes.

We suggest that the long-time evolution of the SH signal is related to the motion of fluxons induced by the variation of the DC magnetic field. Our results point out that the time evolution of the SH signal is characterized by two different regimes. In the time scale of seconds the time evolution of the SH signal is well described by an exponential law. A similar behaviour has been detected in measurements of relaxation of the DC magnetization in multifilamentary Ag-sheathed BSCCO(2223) tapes [17]. The authors, after sweeping the magnetic field at a rate of 33 Oe/s, have detected a fast decay of the DC magnetization in the first 10 s after the field sweep has been stopped. They account for the experimental results by using the Brandt procedure to investigate diffusive motion of fluxon induced by fast variations of the external magnetic field [18, 19]. The similarity between their results and ours strongly suggest that diffusive motion of fluxons is responsible for the SH signal evolution in the time scale of seconds.

At longer times, we have found a logarithmic decay of the SH signal, characterized by a normalized rate of variation (parameter $D$ of Eq. (2)) of the order of $10^{-2}$. Furthermore, our results show that the relaxation rates of the SH signal are different after positive and negative field variations with $D_{H\downarrow} > D_{H\uparrow}$. The dependence of the relaxation rate on the magnetic-field-sweep direction suggests that surface-barrier effects are important in the relaxation of the SH signal. Following the Burlachkov theory [20] on the magnetic relaxation over the surface barrier, it is expected that, for magnetic fields higher than the first-penetration field, after the sample has been exposed to a variation of magnetic field, the magnetic relaxation rates are different for vortex entry and exit. The ratio between the relaxation rate for vortex entry and that for vortex exit depends on the time window in which the relaxation is detected. Rates of variation dependent on the magnetic-field-sweep direction have been observed in the relaxation of the DC magnetization of YBaCuO crystals [21]. At low temperatures and magnetic fields of the order of kOe, the normalized relaxation rates reported in Ref. [21], are of the same order of the ones we obtained in the SH signal decay. Furthermore, the authors show that in the time scale of minutes the relaxation rate measured for vortex exit is greater than that measured for vortex entry, consistently with our results.

The whole process can be figured out as follow. While $H_0$ is varying the surface barrier is absent provided that $H_0 > H_{en}$, for positive field variations, or $H_0 < H_{ex}$, for negative field variations; here $H_{en}$ and $H_{ex}$ are the threshold fields defined by Clem [22] for vortex entry and exit, respectively. During the field sweep the fluxons that cannot follow the field variations are accumulating near the surface. As soon as the field sweep is stopped, a diffusive motion of fluxons sets in; the process ends when the flux density reaches the appropriate value for the critical state. This process may account for the exponential decay of the SH signal in the time scale of seconds. At the same time, relaxation processes take place with different rates for decreasing and increasing field sweeps. At low temperatures, the bulk pinning dominates over the surface barrier, so, in the time window we explore, the flux relaxation is determined by the surface barrier, giving rise to relaxation rates of the SH signal that are different for negative and positive field sweeps.

In conclusion we have reported experimental results of relaxation of the microwave second-order response of YBaCuO powder in different time windows. The sample has been exposed to a DC magnetic field sweep up to a fixed value and the time evolution of the SH signal has been measured for about 300 s, starting from the instant in which the field sweep has been stopped. We have shown that the transient effects have different peculiarities in the different time scales investigated. In the time scale of milliseconds the transient effects in the SH signal of samples in the critical state can be ascribed to processes of fluxon redistribution induced by the switching on/off of the mw pulse. At longer times, the time evolution of the signal can be ascribed to motion of fluxons induced by the variation of the DC magnetic field. After the DC-magnetic-field sweep is stopped, two different processes take place: diffusive motion of fluxons determines the response in the first seconds, magnetic relaxation over the surface barrier determines the response in the time scale of minutes. A further characteristic of the SH emission is the presence of hysteretic behaviour in the field dependence of the SH signal intensity. We have highlighted a strong correlation between the hysteretic behaviour and the transient effects: in the ranges of fields and temperatures in which the magnetic hysteresis is missing, a steady SH response has been detected.

**References**

[1] T. B. Samoilova, Supercond. Sci. Technol. **8** (1995) 259, and references therein.

[2] Q. H. Lam and C. D. Jeffries, Phys. Rev. **B 39** (1989) 4772.

[3] L. Ji, R. H. Sohn, G. C. Spalding, C. J. Lobb, and M. Tinkham, Phys. Rev. **B 40** (1989) 10936.

[4] K. H. Muller, J. C. MacFarlane, and R. Driver, Physica **C 158** (1989) 69.

[5] I. Ciccarello, M. Guccione, and M. Li Vigni, Physica **C 161** (1989) 39.

[6] I. Ciccarello, C. Fazio, M. Guccione, and M. Li Vigni, Physica **C 159** (1989) 769.

[7] M. R. Trunin and G. I. Leviev, J. Phys. III France **2** (1992) 355.

[8] I. Ciccarello, C. Fazio, M. Guccione, M. Li Vigni, and M. R. Trunin, Phys. Rev. **B 49** (1994) 6280

[9] M. A. Golosovsky, H. J. Snortland and M. R. Beasley, Phys. Rev. **B 51**, (1995) 6462.

[10] G. Hampel, B. Batlogg, K. Krishana, N. P. Ong, W. Prusselt, H. Kinder and A. C. Anderson, Appl. Phys. Lett. **71** (1997) 3904.

[11] A. Agliolo Gallitto and M. Li Vigni, Physica **C 305** (1998) 75.

[12] H. Xin, B. E. Oates, G. Dressellhaus and M. S. Dressellhaus, J. Supercond. **14** (2001) 637.

[13] M. Li Vigni, A. Agliolo Gallitto and M. Guccione, Europhys. Lett. **51** (2000) 571.

[14] A. Agliolo Gallitto, M. Li Vigni and D. Scalisi, Physica **C 369** (2002) 245.

[15] A. Agliolo Gallitto, M. Li Vigni and D. Scalisi, Physica **C 377** (2002) 171.

[16] C. P. Bean, Rev. Mod. Phys. **36** (1964) 31.

[17] D. Zola, M. Polichetti and S. Pace, Int. J. Mod. Phys. **14** (2000) 25.

[18] H. E. Brandt, Z. Phys. **B 80** (1990) 167.

[19] H. E. Brandt, Phys. Rev. Lett. **71** (1993) 2821.

[20] L. Burlachkov, Phys. Rev. **B 47** (1993) 8056.

[21]  S. T. Weir, W. J. Nellis, Y. Dalichaouch, B. W. Lee, M. B. Maple, J. Z. Liu and R. N. Shelton, Phys. Rev. **B 43** (1991) 3034.

[22]  J. R. Clem, in *Proceedings of 13$^{th}$ Conference on Low Temperature Physics*, edited by K. D. Timmerhaus, W. J. O'Sullivan, and E. F. Hammel (Plenum, New York, 1974), vol. 3, p.102.

**Figure captions**

Fig. 1  Time evolution of the SH signal in a sample of YBaCuO powder during the first four pulses of the mw field, switched on few seconds after the DC magnetic field has reached the value of 3 kOe, at increasing (circles) and decreasing (squares) fields. $T = 4.2$ K; input peak power $\approx 30$ dBm.

Fig. 2  Time evolution of the SH signal during the very first mw pulse, after the field has reached the value of $H_0 = 3$ kOe at decreasing fields. $T = 4.2$ K; input peak power $\approx 33, 30, 27$ and $24$ dBm, as displayed from the top.

Fig. 3  SH signal intensity as a function of the DC magnetic field. The signal has been detected while the field has been varying from $\sim 300$ Oe to $\sim 10$ kOe, and back, at a rate of $30$ Oe/s. The arrows indicate the path followed during measurements. $T = 4.2$ K; input peak power $\approx 33$ dBm.

Fig. 4  SH signal intensity as a function of the time elapsed from the instant in which the DC magnetic field sweep has been stopped. $T = 4.2$ K; input peak power $\approx 30$ dBm, magnetic-field-sweep rate $\approx 30$ Oe/s. Lines are the best-fit curves obtained using Eqs. (1) and (2) as explained in the text.

Fig. 5  Field dependence of the best-fit parameters $\tau$ and $D$, defined by Eqs. (1) and (2), respectively. $T = 4.2$ K; input peak power $\approx 30$ dBm, magnetic-field-sweep rate $\approx 30$ Oe/s. The lines are leads for eyes.

Fig. 6  Values of the parameters $\tau$ and $D$, deduced by fitting the data using Eqs. (1) and (2), as a function of the magnetic-field-sweeping rate. $H_0 = 3$ kOe; $T = 4.2$ K; input peak power $\approx 30$ dBm.

Fig. 7  Time evolution of the SH signal obtained by the following procedure. The DC magnetic field has been increased from 0 up to 9 kOe, at a rate of 30 Oe/s, in the absence of the mw field; then we have switched on the mw field after a delay time $\Delta t$ from the instant in which the DC field variation had been stopped.

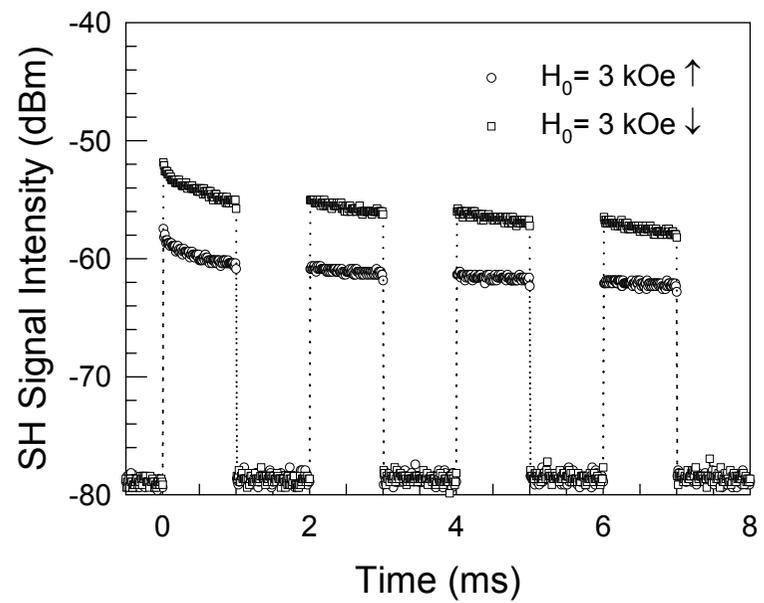

Fig. 1

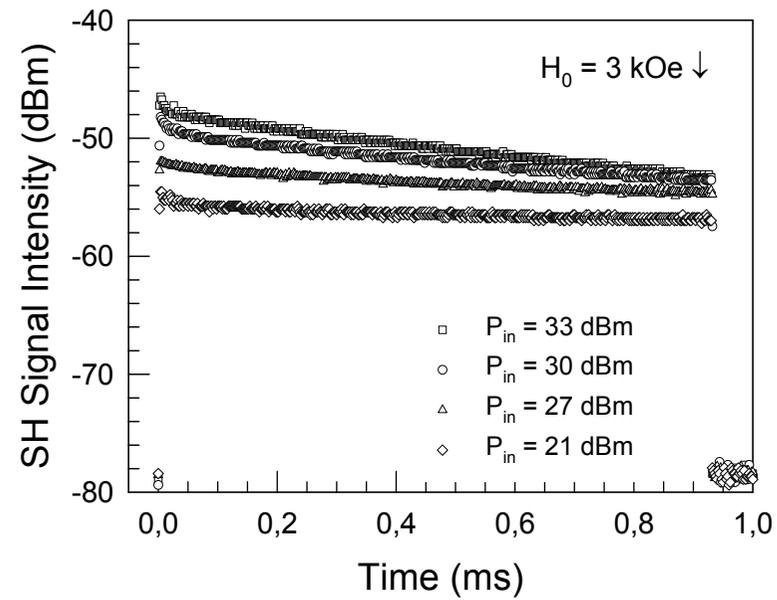

Fig. 2

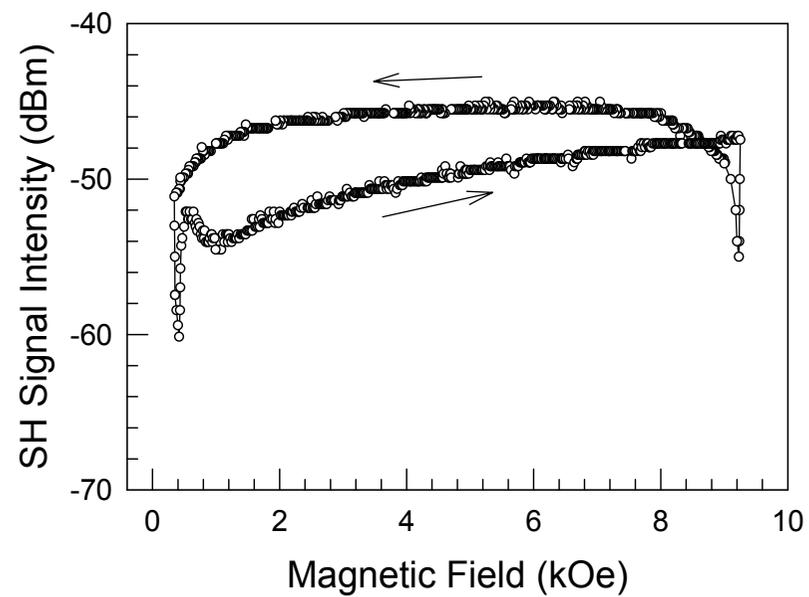

Fig. 3

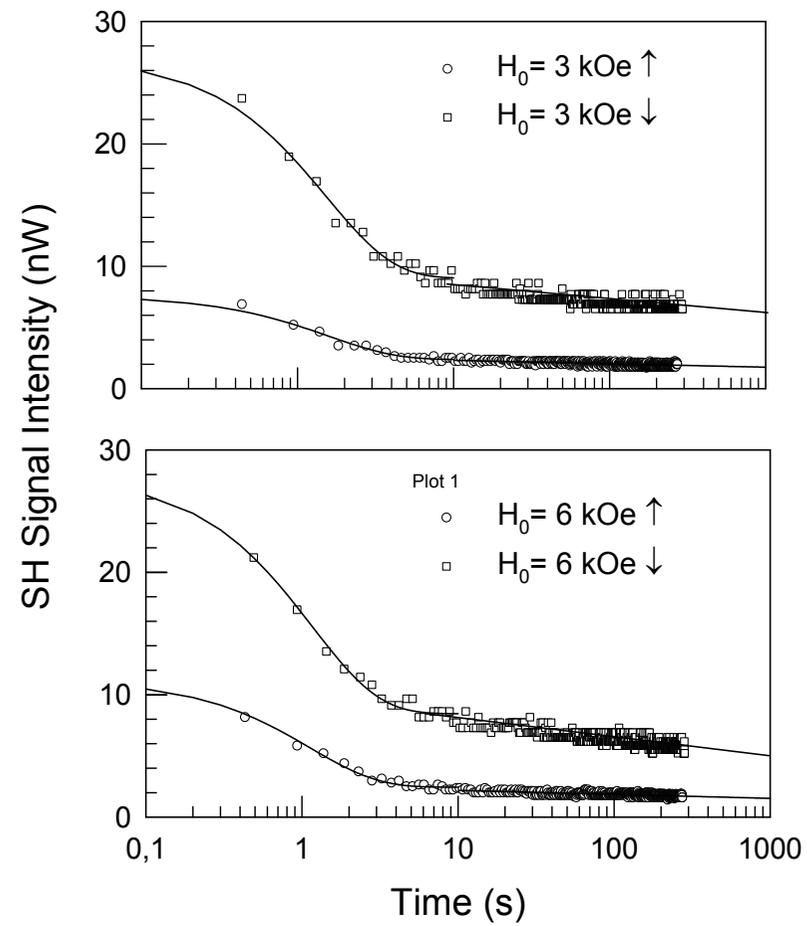

Fig. 4

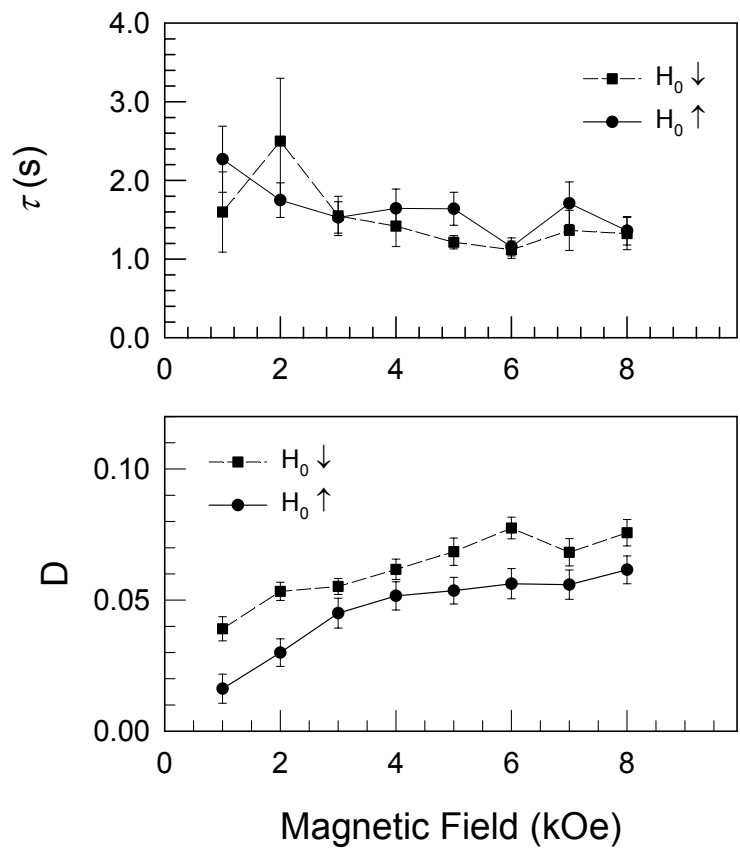

Fig. 5

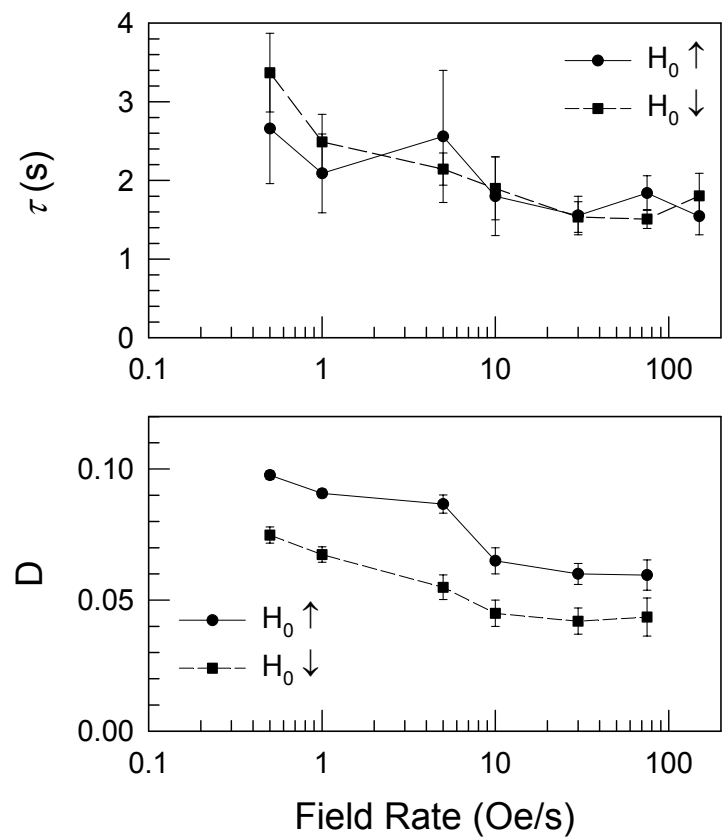

Fig. 6

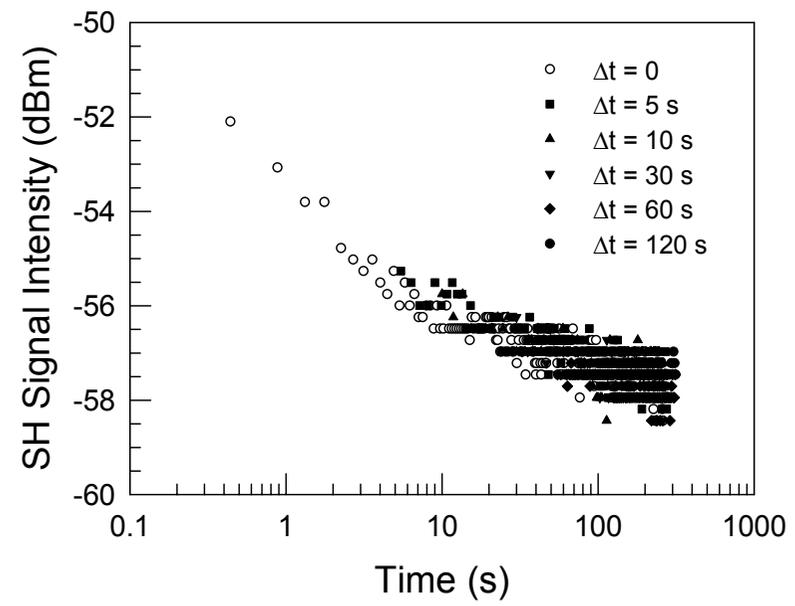

Fig. 7